# Gender differences in research performance within and between countries: Italy vs Norway[1]

**Authors:** Giovanni Abramo[1]\*, Dag W. Aksnes[2], Ciriaco Andrea D'Angelo[3,1]

**Affiliations:**

[1] Laboratory for Studies in Research Evaluation, Institute for System Analysis and Computer Science (IASI-CNR). National Research Council, Rome, Italy
ORCID: 0000-0003-0731-3635 - giovanni.abramo@uniroma2.it

[2] Nordic Institute for Studies in Innovation, Research and Education, Oslo, Norway
ORCID: 0000-0002-1519-195X - dag.w.aksnes@nifu.no

[3] University of Rome 'Tor Vergata', Dept of Engineering and Management, Rome, Italy
ORCID: 0000-0002-6977-6611 - dangelo@dii.uniroma2.it

*\* corresponding author*

**Contributorship**: The authors contributed equally to this work.

**Abstract**
In this study, the scientific performance of Italian and Norwegian university professors is analysed using bibliometric indicators. The study is based on over 36,000 individuals and their publication output during the period 2011–2015. Applying a multidimensional indicator in which several aspects of the research performance are captured, we find large differences in the performance of men and women. These gender differences are evident across all analysed levels, such as country, field, and academic position. However, most of the gender differences can be explained by the tails of the distributions—in particular, there is a much higher proportion of men among the top 10% performing scientists. For the remaining 90% of the population, the gender differences are practically non-existent. The results of the two countries, which differ in terms of the societal role of women, are contrasting. Further, we discuss possible biases that are intrinsic in quantitative performance indicators, which might disfavour female researchers.

---

# 1. Introduction

While the overall representation of women in research is increasing, inequality remains (European Commission, 2019). Women are underrepresented in the research systems of most countries. Only three out of thirty OECD nations—namely, Iceland, Latvia, and Lithuania—have a higher proportion of female researchers than male researchers in higher education (OECD, 2020). In the EU-28 higher education sector, women represent 48% of doctoral students and graduates, 46% of assistant professors, 40% of associate professors, 24% of full professors (15% in STEM), and 22% of heads of institutions (European Commission, 2019). In addition to gender inequalities at a vertical academic rank level, there is also a horizontal segregation where women have a greater presence in certain fields than others (European Commission, 2019). Further, studies have revealed that women progress at a slower rate through academic ranks, tend not to attain important leadership roles, and earn less than men in comparable positions (Wright et al., 2003; McGuire, Bergen, & Polan, 2004; Bilimoria & Liang, 2011; Rotbart, McMillen, Taussig, & Daniels, 2012; Peterson, 2016). The question that naturally arises is which factors, or combination of, could be the cause of this underrepresentation. Broadly speaking, these factors include i) demographic cohort factors related the fact that the current professors have been recruited from cohorts where the female proportion was much lower than today (Stewart, Ornstein, & Drakich, 2009); ii) gender discrimination in the recruitment and advancement processes as well as in the workplace (Regner et al., 2019); iii) self-selection (personal choices)—for example, lower interest among women in pursuing an academic career (Barrett & Barrett, 2011); iv) a competitive meritocratic career system in which women, for various professional and personal reasons, succeed to a lower extent than men (Nielsen, 2016a).

The phenomenon of gender disparities and the underlying causes has been widely studied in the literature from different perspectives—political, economic, sociological, psycho-cognitive, and scientometric. Our investigation belongs to the stream of scientific performance analyses from a scientometric perspective.

Decision-makers increasingly view the investigation of this gap as important for at least two reasons. First, a few wish to develop policies that could eliminate or limit the causes impeding women from achieving their full research performance. Second, while waiting for such policies to take effect, a few are concerned that there must in any case be equitable performance evaluation for male and female researchers, controlling for the factors that could cause a disadvantage. Bibliometrics offers a potent instrument towards providing solutions on both fronts. In fact, bibliometrics has developed different analytical methods and performance indicators that can evidence aspects where there are still gaps between male and female researchers, facilitating the diagnosis of causes. We believe that the extent of the gender gap and the relative relevance of its determinants are country-specific and recommend caution in generalizing findings.

In this study, we present a comparison of the gender representation and scientific performance in the higher education systems of Italy and Norway. The main purpose is to analyse how men and women differ in terms of their scientific performance measured by bibliometric indicators. We address the following questions: What gender patterns can be observed across variables such as country, field, and academic position? Do gender differences occur in the same size among top performing scientists? We also assess the degree of alignment of female representation with female research performance as compared to that of men. The possible reasons for explaining the patterns are considered in a research policy context. Finally, we discuss the possible bias intrinsic in quantitative performance indicators which might disfavour female researchers and provide possible solutions. Given the vast literature on the topic, it is important to establish the value that this research contributes to the existing knowledge on the subject.

First, we believe that the two countries are interesting case studies because of their diversity in terms of gender equality and women rights (Mason, Wolfinger, Goulden, 2013; World Economic Forum, 2020). Norway is generally considered as an exemplary country in this respect. According to the Global Gender Gap Report (World Economic Forum, 2020), Norway ranks second among 153 countries. In the academic system, Norway does not hold a similar top position, with female proportions slightly above the EU-28 average for both PhD graduates and full professors (European Commission, 2019). Italy, on the other hand, ranks as number 76 in the Global Gender Gap ranking, while the female representation in the population of PhD graduates and full professors is close to the EU average.

Compared with Norway, in Italy, women's relative share of involvement in family responsibilities—mainly care for children as well as for parents and parents-in-law—is assumedly more extensive. In Norway, over 90 percent of the children go to kindergarten (Statistics Norway, 2020), thereby allowing parents to be occupied full-time, while this proportion is 24 percent for Italy (Istat, 2019). Further, the percentage of population aged 65 and older is 17 in Norway and 24 in Italy (Population Reference Bureau, PRB, 2020). According to a recent survey, the majority of Italians totally agreed with the statement: 'The most important role of a woman is to take care of her home and family' (European Union, EU, 2017).

Second, the value of this study lies in the method employed to measure and compare research performance. We apply an indicator of performance which accounts not only for outputs but for inputs as well and combines the number of publications with their individual impact. Therefore, it is a multidimensional indicator that encompasses the important aspects of the research performance, while other indicators applied thus far in the literature tend to focus on a single dimension and disregard inputs. Further, we classify authors in fine-grained research fields and compare individual performance within fields to avoid bias due to the different publication patterns across fields. This enables us to observe also whether and to what extent performance gaps differ across fields.

Our analysis does not bear the limits of inferential approaches either, because we observe the entire academic population in the higher education institutions of the two countries and the entire relevant outputs as indexed in the Web of Science (WoS).

The remainder of the paper is structured in the following manner. In the next section, we present a brief review of relevant literature on gender differences in research performance. In Section 3, we present the construction of the data set. In Section 4, we illustrate the method to assess and compare research performance. In Section 5, we present the results of the assessment. In Section 6, we conclude the work with the authors' considerations.

**2. The gender gap in research performance and its determinants**

A consistent finding in the literature is that male researchers generally do indeed publish more than women. This pattern has been revealed across numerous countries, fields, and time periods (Fox, 1983; Cole & Zuckerman, 1984; Long, 1987, 1992; Xie & Shauman, 1998, 2004; Abramo, D'Angelo, & Caprasecca, 2009a; van Arensbergen, van der Weijden, & van den Besselaar, 2012; Sugimoto, Lariviere, Ni, Gingras, & Cronin, 2013; Elsevier, 2020). The gender gap in research performance has been defined as the "productivity puzzle" (Cole and Zuckerman, 1984), although science sociologists retain that when accounting for the role of women in society, the puzzle would vanish. Nevertheless, recent results suggest that the gap is declining, and a large-scale longitudinal study indicated that male and female researchers published equal numbers of papers per year and attributed gender gaps to the shorter career durations of women (Huang et al., 2020) and a larger proportion of women in teaching-intensive and part-time positions (Eagly, 2020). Moreover, there are also studies that reveal a significant number of scientific fields where women outperform men (Abramo, D'Angelo, & Caprasecca, 2009b; Sugimoto, Lariviere, Ni, Gingras, & Cronin, 2013).

Delving into the issue, it becomes evident that the gap occurs in the tails of the performance distribution. Women are more concentrated among very low performers (Lemoine, 1992; Alonso-Arroyo et al., 2007), while the opposite occurs among top performers (Bordons, Morillo, Fernández, & Gómez, 2003; Abramo, D'Angelo, & Caprasecca, 2009b). Moreover, female top scientists are less successful than men in maintaining their stardom over time (Abramo, D'Angelo, & Soldatenkova, 2017).

Men still predominate in the prestigious first- and last-author positions of the byline, and women tend to publish fewer single-authored, first-authored, and last-authored articles than men (Sugimoto, Lariviere, Ni, Gingras, & Cronin, 2013; West, Jacquet, King, Correll, & Bergstrom, 2013). It also appears that women tend to self-cite less than men (Nielsen, 2016b; King, Bergstrom, Correll, Jacquet, & West, 2017), which is a piece of evidence recently challenged by Azoulay and Lynn (2020). Further, when it comes to intellectual property, it has been shown that women patent at approximately 40% of the rate of men (Ding, Murray, & Stuart, 2006).

Findings regarding the average impact of publications diverge. According to some, women perform better (Symonds, Gemmell, Braisher, Gorringe, & Elgar, 2006; Duch et al., 2012), while others demonstrate the opposite (Hunter & Leahey, 2010; Larivière, Vignola-Gagné, Villeneuve, Gelinas, & Gingras, 2011; Aksnes, Rorstad, Piro, & Sivertsen, 2011) or no statistically significant gender difference (Andersen, Schneider, Jagsi, & Nielsen, 2019; Bordons, Morillo, Fernandez, & Gomez, 2003). A very recent study investigated the gender gap in citation impact for six million articles (1996–2018) from Australia, Canada, Ireland, Jamaica, New Zealand, UK, and the USA. Apart from the USA, where differences were unnoticeable, in all other countries, women performed slightly better (Thelwall, 2020).

Various studies have investigated the reasons for the gender differential. In particular, psycho-cognitive studies have explored verbal (Hyde & Linn, 1988), spatial (Linn & Petersen, 1985; Voyer, Voyer, & Bryden, 1995), and mathematical abilities (Hyde, Fennema, & Lamon, 1990; Nowell & Hedges, 1995, 1998). Instead, a few science sociologists have attempted to address the issue from a sociological perspective. When identifying the key drivers of performance, these scholars have remarked the importance of marriage status and stage of the family life cycle. Indeed, differing forms of marriage conduct (Fox, 2005) and the presence of school-aged children appear to negatively affect research productivity (Kyvik & Teigen, 1996; Stacks, 2004). Ley and Hamilton (2008) offer a more direct proof of the effects from maternity by documenting the number of qualified women scientists who stop applying for NIH grants during their late postdoctoral and early faculty years. The inability to obtain projects in the early career stages then tends to become an obstacle for future promotions and, thus, in the further access to funds.

The greater difficulty faced by women in balancing research activities and family duties affects their relative performance (Stack, 2004). However, while the presence of children, particularly those of pre-school age, imply shorter times to be devoted to research, particularly for women, it also constitutes an incentive and motivational driver towards higher time efficiency (Prpić, 2002; Fox, 2005).

The level of research specialization also has a positive relationship with productivity (Abramo, D'Angelo, & Di Costa, 2019), which could explain a part of the negative gap for women. However, findings are not aligned, while Leahey (2006) states that women are generally less specialized than their male colleagues, Abramo, D'Angelo, and Di Costa (2018) indicate that 'there are no gender differences in the propensity to focus or diversify individual research, and urge caution in identifying research diversification as a co-determinant of the gender productivity gap between men and women'.

It has been verified that research collaborations have a positive correlation with scientific performance (Dundar & Lewis, 1998; Lee & Bozeman, 2005; Abramo, D'Angelo, & Di Costa, 2009), particularly collaborations at the international level (Van Raan, 1998; Martin-Sempere, Rey-Rocha, & Garzon-Garcia, 2002; Barjak & Robinson, 2008; Schmoch & Schubert, 2008).

Over the years, numerous studies have investigated whether there are gender differences in collaboration. These studies address collaboration more generally and address international collaboration more specifically (collaboration with authors in other nations). In general, women tend to have more restricted collaboration networks than men (Kyvik & Teigen, 1996; Larivière, Vignola-Gagné, Villeneuve, Gelinas, & Gingras, 2011; Badar, Hite, Badir, 2013; Araújo & Fontainha, 2017), particularly in the initial years of their career (McDowell & Smith, 1992; McDowell, Larry, Singell, & Stater, 2006).

In Italy, female researchers indicate a greater capacity to engage in domestic collaboration, both intramural and extramural, but not in international collaboration, where there is still a gap in comparison to male colleagues (Abramo, D'Angelo, & Murgia, 2013a), probably partially due to motivations against travelling in consideration of family roles. The same occurs in Poland (Kwiek & Roszka, 2020). On the contrary, in Norway, when considering individual variables, virtually no gender differences in the propensity to collaborate internationally were reported (Aksnes, Piro, Rørstad, 2019). Because it is known that internationally co-authored publications are more highly cited than domestic ones (Adams, 2013; Kumar, Rohani, & Ratnavelu, 2014), and this is definitely so in Italy (Abramo, D'Angelo, & Murgia, 2017), one might expect—all other things being equal—different gender performance gaps within and between the two countries. The gender differences in collaboration could partially be determined by gender homophily, whereby scientists prefer to collaborate more with those of the same gender (McDowell & Smith, 1992).

According to Ceci and Williams (2011), differential gender outcomes result exclusively from differences in resources, which might be a consequence of gender discrimination in the workplace. A number of studies have investigated the occurrence of gender discrimination in the recruitment and advancement process, a phenomenon that is inevitably country-specific (Rossiter, 1993; Fuchs, Von Stebut, & Allmendinger, 2001; Wright et al., 2003; Van den Brink, Benschop, & Jansen, 2010; Moss-Racusin, Dovidio,

Brescoll, Graham, & Handelsman, 2012; Abramo, D'Angelo, & Rosati, 2015, 2016b; Zinovyeva & Bagues, 2015).

The question of gender differences in research performance has always been highly sensitive, requiring maximum attention even in discussing issues. It is very likely that it is more the combination of different factors that determines the under-representation of women in research systems and the gender gap in performance, with the weights being variable among countries, institutions, and fields.

## 3. Data

We observe the research activity of Italian and Norwegian professors for the period 2011–2015. We extract data on Italian professors from the database on university personnel maintained by the Italian Ministry of Universities and Research (MIUR, 2020). For each professor, this database provides information on the names and surnames, gender, affiliation, field classification, and academic rank. On 31/12/2015, this data source lists 54,819 professors working at Italian universities; out of these 20,213 (36.9%) are women. We extract Norwegian data from a similar database, the Norwegian Research Personnel Register (providing the official Norwegian R&D statistics, compiled by the Nordic Institute for Studies in Innovation, Research, and Education—NIFU).

The data set used to assess Italian output is extracted from the Italian Observatory of Public Research (ORP), a bibliometric database disambiguated by the author, developed and maintained by Abramo and D'Angelo, and derived under license from the Clarivate Analytics Web of Science (WoS) Core Collection.

Data on publication output of the Norwegian professors is based on a bibliographic database called Cristin (Current Research Information System in Norway), which is a common documentation system for all institutions in the higher education sector, research institutes, and hospitals in Norway. For the period 2011–2015, the database contains data of the publication output of 42,053 individuals (encompassing higher education institutions, hospitals, and institutes), of which 18,673 (44.4%) are women.

For reasons of significance, the analysis is limited to those professors who held formal faculty positions for at least three years over the period 2011–2015. Further, the data set is limited to individuals with at least one publication during the time period (the Norwegian database does not index professors without publications).

To pursue distortion-free comparative assessment (see below), we classify each professor in one and only one WoS subject category, SC. In doing so, we assigned to each publication the SC or SCs of the hosting journal and classified each professor to the most recurrent SC in their publication portfolio. We refer the reader to Abramo, Aksnes, and D'Angelo (2020a) for more details on the procedure.

For reasons of significance, we exclude from the analyses professors in SCs belonging to arts and humanities and a few SCs of the social sciences, where the coverage of WoS has the largest limitations (Hicks, 1999; Larivière, Archambault, Gingras, Vignola-Gagné, 2006; Aksnes & Sivertsen, 2019).

The various SCs differ considerably in size. In order to avoid possible random fluctuations in performance due to a low number of observations, we further exclude those SCs that do not meet the requirements of including i) at least 10 professors in total (summing up Italian and Norwegian ones) and ii) both genders in each country.

The final data set consists of 32,041 Italian and 4,214 Norwegian professors, falling in 145 SCs. Their distribution is presented in Table 1, per academic rank and discipline—that is, clusters of SCs according a pattern previously published on the website of ISI Journal Citation Reports but no longer available on the current Clarivate portal.[2]

**Table 1**
*Data set for analysis*

| Discipline | No. of SCs | Italy | | | | Norway | | | |
| --- | --- | --- | --- | --- | --- | --- | --- | --- | --- |
| | | Tot. professors | Assistant (%) | Associate (%) | Full (%) | Tot. professors | Assistant (%) | Associate (%) | Full (%) |
| Mathematics | 4 | 2065 | 23.0 | 40.6 | 36.4 | 175 | 2.3 | 29.1 | 68.6 |
| Physics | 13 | 2680 | 24.2 | 42.8 | 33.0 | 241 | 10.4 | 21.2 | 68.5 |
| Chemistry | 5 | 1588 | 29.2 | 42.6 | 28.2 | 109 | 15.6 | 29.4 | 55.0 |
| Earth and Space sciences | 10 | 1792 | 29.6 | 41.6 | 28.8 | 410 | 14.9 | 28.5 | 56.6 |
| Biology | 22 | 5315 | 35.2 | 37.5 | 27.2 | 722 | 20.9 | 28.3 | 50.8 |
| Biomedical research | 11 | 3564 | 37.9 | 36.4 | 25.7 | 237 | 19.8 | 30.4 | 49.8 |
| Clinical medicine | 33 | 7463 | 35.8 | 36.0 | 28.1 | 955 | 10.6 | 31.2 | 58.2 |
| Psychology | 6 | 475 | 31.2 | 39.8 | 29.1 | 148 | 2.7 | 42.6 | 54.7 |
| Engineering | 22 | 4809 | 26.6 | 40.2 | 33.2 | 377 | 4.8 | 26.3 | 69.0 |
| Political and social sciences | 11 | 442 | 20.6 | 41.6 | 37.8 | 438 | 6.2 | 33.1 | 60.7 |
| Economics | 8 | 1848 | 19.0 | 40.4 | 40.6 | 402 | 3.0 | 33.1 | 63.9 |
| *Total* | *145* | *32041* | *30.8* | *38.8* | *30.3* | *4214* | *11.1* | *30.0* | *58.9* |

## 4. Measuring research productivity: The FSS indicator

Aligning with the microeconomic theory of production, we measure professors' performance by their research productivity. Productivity is the quintessential indicator of efficiency in any production system. It is commonly defined as the rate of output per unit of input. It measures how efficiently production inputs are being used. Most bibliometricians measure research productivity by the number of publications in the period under observation. Because publications (output) have different values and resources employed for research are not homogenous across individuals and organizations, a more appropriate definition of productivity in research systems is the value of output per Euro spent in research. Bibliometricians value publications by means of their scholarly impact, which they measure by citations. Because citation behaviour varies according to field, we standardize the citations for each publication with respect to the average of the distribution of citations for all the cited publications indexed in the same year and the same SC.

Further, we account for the fractional contributions of scientists to outputs, which is occasionally further signalled by the position of the authors in the list of authors. Unfortunately, relevant data on the resources used by each researcher are not always available and a number of assumptions are required. We assume that the same resources are available to all professors within the same field and that the hours devoted to educational activities are more or less the same for all professors.

To measure the yearly average research productivity of Italian and Norwegian academics, we use the fractional scientific strength (FSS) indicator, a proxy measure of research productivity. A thorough description of the FSS indicator and the theory underlying it can be found in Abramo and D'Angelo (2014).

---
[2] For the complete data set and the SC clustering, refer to the Supplementary material.

The FSS formula is as given below:

$$FSS = \frac{1}{\left(\frac{w_r}{2} + k\right)} \cdot \frac{1}{t} \sum_{i=1}^{N} \frac{c_i}{\bar{c}} f_i,$$

[1]

where
$w_r$ = average yearly salary of professor
$k$ = average yearly capital available for research to professor
$t$ = number of years of work by the professor in period under observation
$N$ = number of publications by the professor in period under observation
$c_i$ = citations received by publication $i$, until 31 October 2018 (citation time window ranges from 3 to 7 years).
$\bar{c}$ = average of distribution of citations received for all WoS cited publications in the same year and SC of publication $i$
$f_i$ = fractional contribution of professor to publication $i$.

With regard to the input, the underlying assumption is that labour and capital equally contribute to production, but we halved labour costs assuming that 50 percent of professors' time is allocated to activities other than research. Sources of input data and details on the measurement can be found in Abramo, Aksnes, and D'Angelo (2020a). In particular, in both countries, there is no gender gap in wages for academic positions, and we assume that $k$ is the same for men and women. We are also assuming a correspondence of academic ranks in Norway and Italy.

The fractional contribution equals the inverse of the number of authors in those fields where the practice is to place the authors in simple alphabetical order, but it assumes different weights in other cases. Specifically, for Biology, Biomedical research, and Clinical medicine, widespread practice in both Italy and Norway is for the authors to indicate the various contributions to the published research by the order of the names in the bylines. Thus, for the above disciplines, we ascribe different weights to each co-author according to their position in the list of authors and the character of the co-authorship (intramural or extramural). If the publication is the outcome of an exclusively intramural collaboration (only one affiliation in the address list), 40% is attributed to both the first and last authors and the remaining 20% is divided among all other authors. In contrast, if the publication address list indicates extramural collaborations, 30% is attributed to both first and last authors; 15% to both second and last but one author; and the remaining 10% is divided among all other authors. The weighting values were assigned following advice from senior Italian professors in the life sciences and can be changed to suit different practices in other national contexts.

The FSS of professors belonging to different SCs cannot be compared directly. In fact, i) scientists' intensity of publication varies remarkably across fields in general (Sandström & Sandström, 2009; Lillquist & Green, 2010; Sorzano, Vargas, Caffarena-Fernández, & Iriarte, 2014) and in both countries in particular (Piro, Aksnes & Rørstad, 2013; D'Angelo & Abramo, 2015); ii) the intensity of collaboration—that is, the average number of co-authors per publication—also varies across fields (Glanzel & Schubert, 2004; Yoshikane & Kageura, 2004; Abramo, D'Angelo, & Murgia, 2013b). To avoid distortions then, the performance rankings of professors are constructed at the SC level.

For comparisons at higher levels of aggregation—that is, discipline and overall—as suggested by Abramo, Cicero, and D'Angelo (2012), we normalize FSS scores to the average score of all professors of the same SC, but those with nil score.

At the aggregate level then, the yearly productivity $FSS_A$ for the aggregate unit $A$ is

$$FSS_A = \frac{1}{RS}\sum_{j=1}^{RS}\frac{FSS_j}{\overline{FSS}},$$

[2]

where
$RS$ = number of professors in the unit in the observed period;
$FSS_j$ = productivity of professor $j$ in the unit;
$\overline{FSS}$ = average productivity of all productive professors under observation in the same SC of professor $j$.

Note that the scaling does not refer to world distributions. Since we are comparing Italy and Norway, the 'average' used to rescale original distributions is calculated by collapsing Italian and Norwegian performance distributions only. To exemplify, an FSS score of 1.10 means that the professor's performance is 10% above average of all Italian and Norwegian professors in his or her own SC. In the following tables, figures, and text, all FSS scores are normalized but we maintain the same denomination for FSS.

## 5. Results

In this section, first we analyse the gender representation by academic rank and discipline in the Italian and Norwegian academic systems. Then, we compare research performance by gender, within and between countries. Finally, we analyse the degree of alignment between representation and performance, across academic ranks.

### 5.1 Gender representation

Overall, women represent around one-third of the total academic staff under observation (last line of Table 2), both in Italy (33.8%) and Norway (33.9%). The variability in gender representation among academic ranks differs between the two countries. In Italy, women represent 47.2% of the total assistant professors, 35.2% of associate professors, and 18.3% of full professors; in Norway, they represent 41.5%, 46.5%, and 26.1%, respectively. In Norway, women representation in higher academic ranks is higher than that in Italy.

Gender data disaggregated by discipline indicate a considerable differentiation among fields (Figure 1). In Italy, the gender composition has the largest imbalance in Physics and Engineering, where women account for less than 20% of the total professors. In Norway, the situation is very similar, but the same is true in Mathematics, where only 13.7% of the total number of professors are women, against 36.7% in Italy. Moreover, in Earth and space sciences, the incidence of women is lower than the average overall value in both countries (29.7% in Italy; 25.9% in Norway). The comparison between the two countries indicates that in Chemistry, the proportion of women in Italy (40.9%) is higher than average, while in Norway it is lower (27.5%). In Italy, the proportion of women is highest in psychology (62.1%) and in the life sciences; in biology the proportion is 47.4%, and in biomedical research it is 45.0%. In Norway, the highest proportion of women is found in clinical medicine (48.7%), political and social sciences (46.8%), and psychology (43.9%). It must be noted that in Italy, the proportion of female professors in psychology is higher than that of males, which is not the case in Norway.

With regard to the proportion of women across academic ranks (Table 2), the underrepresentation of women in Norway is particularly strong in mathematics, whereby, as said above, the proportion of women is 13.7% of total professors and only 8.3% of full professors. In Italy, the lowest share among full professors is found in physics (9.9%). In Norway, the proportion of women among full professors is higher than that overall (33.9%) in clinical medicine (36.7%), psychology (44.4), and political and social sciences (43.6%). In Italy, the higher proportion of women is only found in psychology (48.6% vs 33.8% overall).

**Table 2**
*Number of professors and proportion of women in the two countries by discipline and academic rank*

| | Italy | | | | | Norway | | | | |
| | | Proportion of women | | | | | Proportion of women | | | |
| Discipline | Total professors | Overall (%) | Assistant prof. (%) | Associate prof. (%) | Full prof. (%) | Total professors | Overall (%) | Assistant prof. (%) | Associate prof. (%) | Full prof. (%) |
|---|---|---|---|---|---|---|---|---|---|---|
| Mathematics | 2065 | 36.7 | 49.1 | 43.1 | 21.8 | 175 | 13.7 | 25.0 | 25.5 | 8.3 |
| Physics | 2680 | 17.2 | 25.6 | 18.0 | 9.9 | 241 | 14.9 | 24.0 | 19.6 | 12.1 |
| Chemistry | 1588 | 40.9 | 59.3 | 41.4 | 21.2 | 109 | 27.5 | 29.4 | 31.3 | 25.0 |
| Earth and space sciences | 1792 | 29.7 | 39.9 | 33.0 | 14.3 | 410 | 25.9 | 34.4 | 38.5 | 17.2 |
| Biology | 5315 | 47.4 | 61.0 | 50.3 | 26.0 | 722 | 33.2 | 40.4 | 42.6 | 25.1 |
| Biomedical research | 3564 | 45.0 | 60.0 | 44.6 | 23.7 | 237 | 41.4 | 31.9 | 62.5 | 32.2 |
| Clinical medicine | 7463 | 30.7 | 43.2 | 30.3 | 15.4 | 955 | 48.7 | 61.4 | 66.8 | 36.7 |
| Psychology | 475 | 62.1 | 70.3 | 65.6 | 48.6 | 148 | 43.9 | 0.0 | 46.0 | 44.4 |
| Engineering | 4809 | 19.2 | 27.5 | 20.4 | 11.2 | 377 | 17.5 | 27.8 | 23.2 | 14.6 |
| Political and social sciences | 442 | 42.1 | 60.4 | 42.9 | 31.1 | 438 | 46.8 | 48.1 | 52.4 | 43.6 |
| Economics | 1848 | 32.0 | 44.9 | 38.7 | 19.3 | 402 | 23.6 | 41.7 | 38.3 | 15.2 |
| *Total* | *32041* | *33.8* | *47.2* | *35.2* | *18.3* | *4214* | *33.9* | *41.5%* | *46.5%* | *26.1%* |

**Figure 1**
*The proportion of female professors by discipline in the two countries (horizontal dashed lines = gender representation at overall level—33.8% for Italy, 33.9% for Norway)*

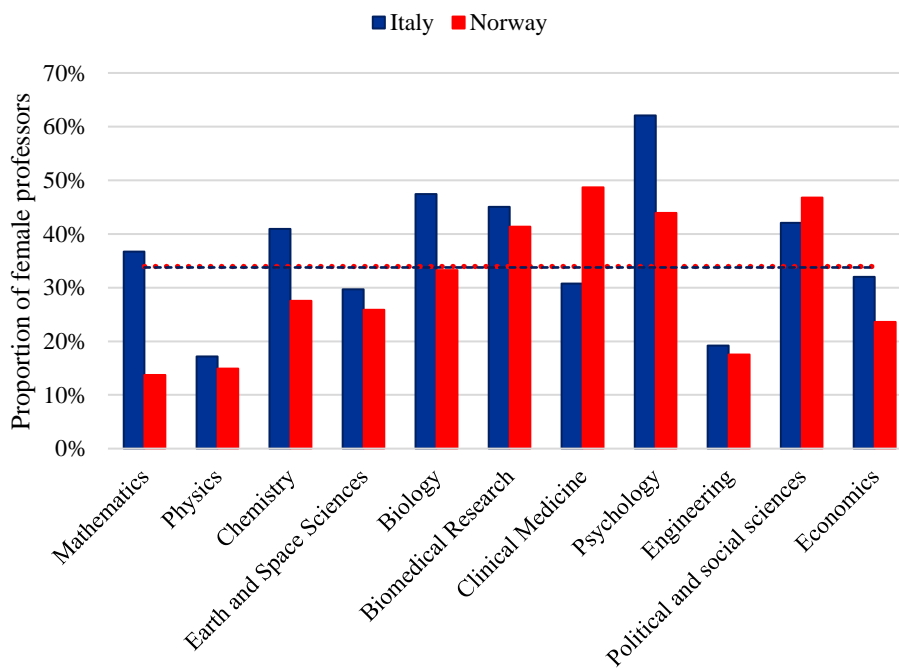

## 5.2 Research performance analysis

In this subsection, we present the analysis of gender differences in performance within and between countries, per SC, discipline, and overall. We remind the reader that the FSS scores presented are all normalized.

In aggregating individual level performances, we must consider that bibliometric data are generally not normally distributed, showing high skewness due to the presence of outliers on the right tail. However, for the evaluation of groups, departments, institutions, and countries, bibliometricians almost always use mean values. Then, we present and discuss the mean values of performance but also other kind of statistics of the relevant distributions (medians, deciles, and quartiles). Specifically, when presenting differences, we utilize nonparametric statistical significant testing.

Overall, men outperform women in both countries (Table 3). In Italy, the average performance of men is 37% higher than that of women (FSS is 1.09 for men vs 0.79 for women). In Norway, it is 32% higher (1.08 vs 0.82). Significant differences are also recorded for the median values: 0.60 vs 0.47 in Italy, and 0.54 vs 0.47 in Norway.

The top male Norwegian performer indicates a normalized FSS (19.61)—that is, 70% higher than the top female (11.33). In Italy, the scores are respectively 74.85 and 44.54.

The difference between Italy and Norway is extremely large for max values, while for other statistics (average, median, standard deviation, etc.) it is rather small. This is the effect of the presence of outliers on the right tail of the relevant distributions—a right tail that is longer for Italy than for Norway.

In general, the distribution of men' performance is more dispersed and skewed than that of women in both countries. The interquartile range of the FSS distribution is 1.12 for Italian men vs 0.81 for Italian women. In Norway, it is 1.16 for men and 0.88 for women.

The comparison between the two countries indicates that the proportion of uncited professors (% with nil FSS) is higher in Norway for both men and women, although uncited professors account for a marginal proportion of the population in both countries. In terms of average FSS, there are small differences across the nations, with a slight superiority of male professors in Italy and of female professors in Norway. Differences recorded for median values are statistically significant only for men.

**Table 3**
*Descriptive statistics of normalized performance (FSS), by gender and country*

|  | Italy | | | Norway | | | Italy vs Norway | |
| --- | --- | --- | --- | --- | --- | --- | --- | --- |
|  | M | F | Δ | M | F | Δ | ΔM | ΔF |
| No. of professors | 21224 | 10817 | 10407 | 2784 | 1430 | 1354 | 18440 | 9387 |
| % with nil FSS | 0.9% | 1.0% | -0.1% | 2.2% | 2.0% | 0.2% | -1.3% | -1.0% |
| Avg. FSS | 1.09 | 0.79 | 0.30 | 1.08 | 0.82 | 0.26 | 0.01 | -0.03 |
| Median FSS | 0.60 | 0.47 | 0.13*** | 0.54 | 0.47 | 0.06*** | 0.07** | 0.01 |
| Q1 | 0.23 | 0.19 | 0.04 | 0.19 | 0.17 | 0.02 | 0.03 | 0.01 |
| IQR† | 1.12 | 0.81 | 0.32 | 1.16 | 0.88 | 0.28 | -0.03 | -0.07 |
| Max FSS | 74.85 | 44.54 | 30.30 | 19.61 | 11.33 | 8.28 | 55.23 | 33.21 |
| FSS st. dev. | 1.68 | 1.10 | 0.58 | 1.55 | 1.11 | 0.44 | 0.13 | -0.01 |
| FSS skewness | 9.23 | 9.04 | 0.19 | 3.83 | 3.87 | -0.04 | 5.40 | 5.17 |

*Two-sample Wilcoxon rank-sum (Mann-Whitney) test. Statistical significance: \*p-value <0.10, \*\*p-value <0.05, \*\*\*p-value <0.01*
† *Interquartile range (Q3–Q1)—that is, the difference between the third and first quartiles.*

In order to unveil where the FSS gender gaps highlighted in Table 3 originate from, we delve into the constituent components of the FSS indicator—namely, number of publications, average citations per paper, and contribution to paper. As in Abramo, Aksnes, and D'Angelo (2020b), we also investigate the average impact measured by the impact factor of hosting journals. Table 4 presents the average and median normalized values of the following aspects:
- Output (O)—that is, average yearly publications.
- Fractional output (FO)—that is, average yearly total contribution to publications.
- Average citations (AC)—that is, average standardized citations per publication.
- Average impact factor (AIF)—that is, average standardized IF per publication.

As for Italy, there occurs a significant gap in favour of men, by O (1.08 vs 0.89 for average values; 0.83 vs 0.71 for medians). The gap increases when accounting for the actual contribution to each publication (1.09 vs 0.84 for average values; 0.82 vs 0.65 for medians), confirming that Italian women tend to collaborate more than men. The gaps in average citations per paper and average IF are negligible (even if statistically significant). In Norway, it is almost the same, with no differences at all in average citation per paper and average IF. Italian men and women publish more than Norwegian ones, but when adopting fractional counting, differences are not significant. With regard to impact, there are no differences in terms of citations, while the average IF of the publications of Italian male and female professors is lower than that for Norwegian ones.

**Table 4**
*Average and median normalized values of FSS constituent components, by gender and country*

|  |  | Italy | | | Norway | | | Italy vs Norway | |
|---|---|---|---|---|---|---|---|---|---|
| Indicator | | M | F | Δ† | M | F | Δ† | ΔM† | ΔF† |
| O | Avg. | 1.08 | 0.89 | 0.19 | 0.96 | 0.75 | 0.21 | 0.12 | 0.14 |
|  | Median | 0.83 | 0.71 | 0.12*** | 0.69 | 0.54 | 0.15*** | 0.14*** | 0.17*** |
| FO | Avg. | 1.09 | 0.84 | 0.24 | 1.05 | 0.82 | 0.23 | 0.04 | 0.02 |
|  | Median | 0.82 | 0.65 | 0.16*** | 0.79 | 0.62 | 0.17*** | 0.03 | 0.04 |
| AC | Avg. | 1.01 | 0.97 | 0.03 | 1.03 | 1.03 | 0.00 | -0.03 | -0.06 |
|  | Median | 0.85 | 0.82 | 0.04*** | 0.83 | 0.86 | -0.03 | 0.02** | -0.04 |
| AIF | Avg. | 1.00 | 0.98 | 0.01 | 1.07 | 1.07 | -0.01 | -0.07 | -0.09 |
|  | Median | 0.96 | 0.94 | 0.01*** | 1.01 | 1.01 | 0.00 | -0.05*** | -0.06*** |

*Two-sample Wilcoxon rank-sum (Mann-Whitney) test. Statistical significance: \*p-value < 0.10, \*\*p-value < 0.05, \*\*\*p-value < 0.01*

† *We show only two decimal digits for M and F values, but for Δ calculation we use the values without approximation.*

Figures 2 and 3 provide an overall picture of the FSS distribution by gender in the two countries. For each performance decile, the graphs indicate the proportion of women (and their complement, of men), compared to the expected value which, we recall, is the gender representation—that is, approximately two men for every woman.

For Italy (Figure 3), the proportion of female professors is higher than expected in all low FSS deciles. From the sixth decile onwards, the proportion of women progressively decreases and, accordingly, the proportion of men increases. Among the top 10% performers, women account only for 22.0% of the population.

**Figure 2**
*Normalized performance (FSS percentile) distribution of Italian professors, by gender (dashed lines = expected values—that is, 33.8% for women, 66.2% for men)*

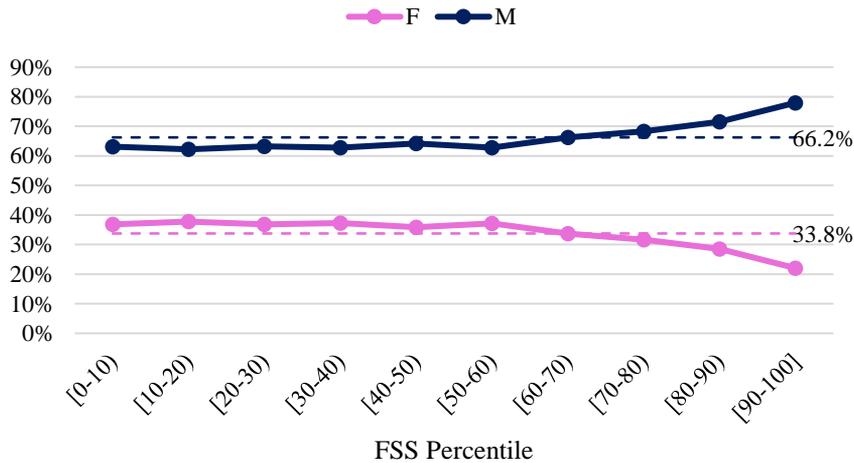

For Norway, the situation is depicted in Figure 3. For the first two lowest deciles, the proportion of women is similar to the expected value. In the next five deciles, there is a slight female over-representation that fades away in the eighth and ninth deciles. The similarity with the Italian case is evident only for the top decile, where the proportion of women drops to 21.3%.

A comparison between the two countries is presented in Figure 4. Data reveal an over-representation of Italian women compared to Norwegians in the first four bottom deciles of performance and the opposite is true in the next three deciles (6–9). Among the top 10% of professors, the proportion of women is similar in the two countries.

**Figure 3**
*Normalized performance (FSS percentile) distribution of Norwegian professors, by gender (dashed lines = expected values—that is, 33.9% for women, 66.1% for men)*

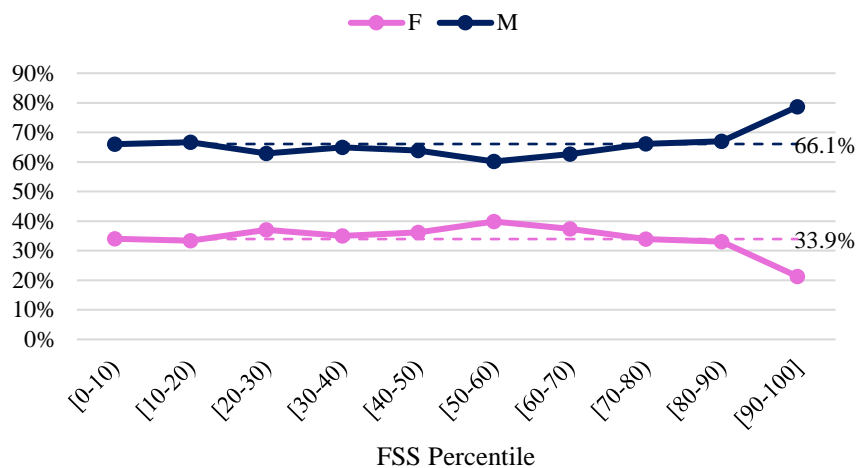

**Figure 4**
*Female normalized performance (FSS) distributions in the two countries (horizontal dashed lines = gender representation)*

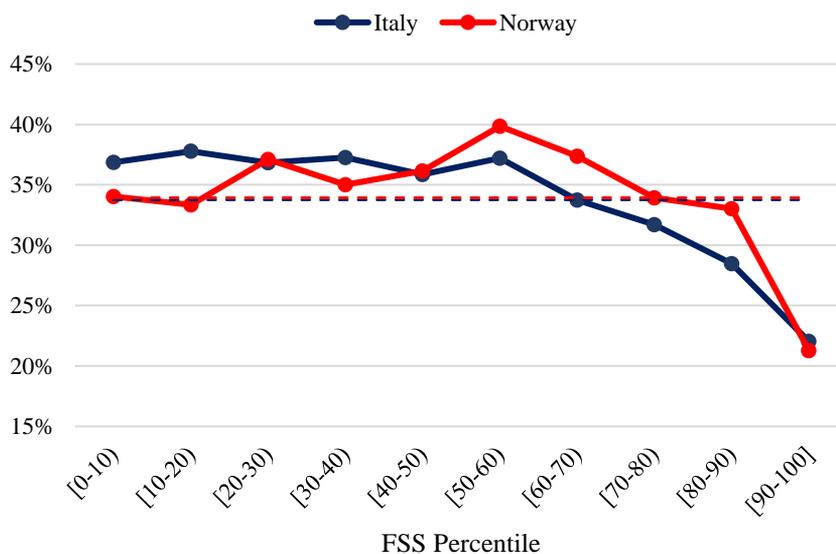

The overlapping of the gender performance distributions of the two countries in the right tail suggests to deepen the analysis within the top quartile. Considering an x-th percentile threshold, in the absence of gender performance gap, we expect to observe the two genders equally represented—that is, exactly x% of total men and x% of total women (allotting minus or plus decimal points, due to ties in the rankings). Table 5 proposes this kind of analysis with four different thresholds, within and between the two countries. Values recorded for Italy indicate a male concentration that is three times higher than female concentration among the top 1% professors (1.5% vs 0.5%). Lowering the threshold to top 5%, the male concentration is double that of women (6.2% vs 3.0%).

Moreover, for the other two thresholds (top 10% and top quartile), the percentage of men is higher than expected in the absence of a gender gap; consequently, the opposite is true for women. As shown in columns 5–7 of Table 5, the pattern for Norway is rather similar. The difference in concentration between the two genders is even greater than that recorded for Italy.

Comparing gender differences "between" the two countries (last two columns), for men, we observe a greater concentration in Norway for all the thresholds. It is the same for women, with the only exception of the 5% threshold with no difference.

**Table 5**
*Proportion of top professors by FSS, by gender and country*

|  | Italy | | | Norway | | | Italy vs Norway | |
|---|---|---|---|---|---|---|---|---|
|  | M | F | Total | M | F | Total | ΔM | ΔF |
| Top 1% | 1.5% | 0.5% | 1.2% | 2.0% | 0.8% | 1.6% | -0.5% | -0.2% |
| Top 5% | 6.2% | 3.0% | 5.1% | 7.0% | 3.0% | 5.6% | -0.8% | 0.0% |
| Top 10% | 11.9% | 6.6% | 10.1% | 12.8% | 6.7% | 10.7% | -0.9% | -0.1% |
| Top 25% | 27.8% | 19.7% | 25.1% | 28.1% | 21.5% | 25.8% | -0.3% | -1.7% |

### 5.2.1 Gender performance differences among research fields

In this section, we analyse the gender performance gap across SCs and disciplines. Table 6 presents that in both countries, men outperform women in all disciplines. In Italy, the average FSS difference is the largest in biomedical research (1.22 for men vs 0.75 for women) and smallest in economics (1.03 vs 0.86). In Norway, the largest gap occurs in mathematics (1.26 vs 0.49), and the smallest in political and social sciences (0.91 vs 0.87). In Norway, engineering stands out, as it is the second-last discipline in terms of female representation (17.5%) and the same in terms of performance gap (-0.09). In other words, although there are relatively few women in this discipline, the performance of men and women does not differ as much as in most other disciplines.

In comparing the two countries, we note that in all disciplines, except mathematics, earth and space sciences, and clinical medicine, Italian men outperform Norwegian men, with biomedical research as top high (1.22 vs 0.98). The opposite is true for women, whereby Norwegian women show higher average performance than Italian in all disciplines but mathematics, physics, biology and economics, with earth and space sciences as a top high (1.09 vs 0.72). The case of mathematics is peculiar, as it is the discipline with the highest performance gap in favour of Norwegian men (0.20) and Italian women (0.28).

**Table 6**
*Gender differences in average normalized performance (FSS), within and between countries*

|  | Italy | | | Norway | | | Italy vs Norway | |
|---|---|---|---|---|---|---|---|---|
| Discipline | M | F | $\Delta^{\dagger}$ | M | F | $\Delta^{\dagger}$ | $\Delta M^{\dagger}$ | $\Delta F^{\dagger}$ |
| Mathematics | 1.06 | 0.77 | 0.29 | 1.26 | 0.49 | 0.77 | -0.20 | 0.28 |
| Physics | 1.04 | 0.82 | 0.22 | 0.98 | 0.69 | 0.29 | 0.06 | 0.12 |
| Chemistry | 1.08 | 0.88 | 0.21 | 1.03 | 0.94 | 0.09 | 0.05 | -0.06 |
| Earth and Space Sciences | 1.02 | 0.72 | 0.30 | 1.34 | 1.09 | 0.25 | -0.32 | -0.36 |
| Biology | 1.15 | 0.85 | 0.30 | 1.04 | 0.66 | 0.38 | 0.11 | 0.19 |
| Biomedical Research | 1.22 | 0.75 | 0.47 | 0.98 | 0.75 | 0.23 | 0.24 | 0.00 |
| Clinical Medicine | 1.10 | 0.73 | 0.37 | 1.18 | 0.87 | 0.32 | -0.08 | -0.13 |

| | | | | | | | |
|---|---|---|---|---|---|---|---|
| Psychology | 1.20 | 0.85 | 0.35 | 1.15 | 0.87 | 0.28 | 0.06 | -0.02 |
| Engineering | 1.04 | 0.80 | 0.24 | 0.91 | 0.82 | 0.09 | 0.13 | -0.02 |
| Political and Social Sciences | 1.01 | 0.77 | 0.24 | 0.91 | 0.87 | 0.05 | 0.09 | -0.10 |
| Economics | 1.03 | 0.86 | 0.18 | 1.03 | 0.73 | 0.30 | 0.00 | 0.12 |

† *We show only two decimal digits for M and F values, but for Δ calculation we use all decimals.*

In Table 7, we repeat the same analysis using the median value of FSS in order to account for the skewness of the FSS distribution. It is confirmed that men outperform women in all disciplines, both in Italy and in Norway. However, the Wilcoxon rank-sum test shows that the performance gap is not always statistically significant, particularly in Norway. In comparing the two countries, we note that the FSS medians of Italian and Norwegian professors are rather close, with a slight predominance (in all disciplines, but Clinical Medicine and Political and social sciences) for Italians. With regard to women, Norwegian women have a higher median in four disciplines, but the differences are statistically significant only in earth and space sciences and in political and social sciences. The opposite is true only in mathematics and biology.

At a lower level of aggregation, Table 8 presents the number of SCs where the average FSS of women is higher than that registered for men, by discipline and country.[3]

In Italy, women outperform men in 23 SCs (i.e. 15.9% of total cases). In Norway, in 49 SCs (33.8%). In Italy, in mathematics and chemistry, there are no SCs in which women outperform men. In Norway, it is so in mathematics only. In Italy, the highest proportions of SCs in a discipline where women outperform men is found in political and social sciences (27.3%); it is the same in Norway, but with much higher proportions (63.6%).

**Table 7**
*Gender differences in median normalized performance (FSS), within and between countries*

| | Italy | | | Norway | | | Italy vs Norway | |
|---|---|---|---|---|---|---|---|---|
| Discipline | M | F | Δ† | M | F | Δ† | ΔM† | ΔF† |
| Mathematics | 0.54 | 0.42 | 0.12*** | 0.52 | 0.33 | 0.19** | 0.02** | 0.10** |
| Physics | 0.63 | 0.54 | 0.09*** | 0.63 | 0.54 | 0.10 | 0.00 | 0.00 |
| Chemistry | 0.66 | 0.60 | 0.05** | 0.63 | 0.36 | 0.28 | 0.02 | 0.25 |
| Earth and Space Sciences | 0.66 | 0.50 | 0.17*** | 0.67 | 0.63 | 0.03 | 0.00** | -0.14*** |
| Biology | 0.67 | 0.53 | 0.14*** | 0.52 | 0.48 | 0.05*** | 0.15 | 0.06* |
| Biomedical Research | 0.57 | 0.43 | 0.15*** | 0.47 | 0.47 | 0.00 | 0.10 | -0.04 |
| Clinical Medicine | 0.53 | 0.41 | 0.12*** | 0.63 | 0.50 | 0.14** | -0.10*** | -0.09 |
| Psychology | 0.75 | 0.57 | 0.19** | 0.56 | 0.38 | 0.18 | 0.19 | 0.19 |
| Engineering | 0.64 | 0.49 | 0.15*** | 0.48 | 0.42 | 0.06 | 0.17 | 0.08 |
| Political and social sciences | 0.44 | 0.34 | 0.10 | 0.46 | 0.41 | 0.05 | -0.02 | -0.07* |
| Economics | 0.52 | 0.45 | 0.07*** | 0.47 | 0.42 | 0.05 | 0.05 | 0.03 |

*Two-sample Wilcoxon rank-sum (Mann-Whitney) test. Statistical significance: \*p-value < 0.10, \*\*p-value < 0.05, \*\*\*p-value < 0.01*

† *We show only two decimal digits for M and F values, but for Δ calculation we use the values without approximation.*

**Table 8**
*Number and proportion of subject categories (SCs) where women outperform men (in terms of mean FSS), by discipline and country*

| | | Italy | Norway |
|---|---|---|---|
| Discipline | No. of SCs | F higher in | F higher in |
| Mathematics | 4 | 0  0.0% | 0  0.0% |

---

[3] Here, the count is referred to differences in averages values only, regardless of their possible statistical significance.

| | | | | | |
|---|---|---|---|---|---|
| Physics | 13 | 2 | 15.4% | 4 | 30.8% |
| Chemistry | 5 | 0 | 0.0% | 2 | 40.0% |
| Earth and Space Sciences | 10 | 2 | 20.0% | 3 | 30.0% |
| Biology | 22 | 4 | 18.2% | 6 | 27.3% |
| Biomedical Research | 11 | 1 | 9.1% | 4 | 36.4% |
| Clinical Medicine | 33 | 5 | 15.2% | 11 | 33.3% |
| Psychology | 6 | 1 | 16.7% | 1 | 16.7% |
| Engineering | 22 | 4 | 18.2% | 9 | 40.9% |
| Political and social sciences | 11 | 3 | 27.3% | 7 | 63.6% |
| Economics | 8 | 1 | 12.5% | 2 | 25.0% |
| Overall | 145 | 23 | 15.9% | 49 | 33.8% |

We delve deeper into the gender performance gaps at SC level, thereby revealing the top ten SCs by performance gap in favour of either gender, in both countries. Table 9 presents the case of Italy. In general, when examining the last column, we note that the upper portion of the table indicates much larger score differences than the lower. Three SCs (physics, nuclear, telecommunications, and transportation) present a peculiarity: despite female representation being relatively low (19 out of 92, 22 out of 157, and 13 out of 79, respectively), their average performance is noticeably higher than that of men.

Table 10 presents the same analysis for Norway. Here too, there are a couple of SCs—namely, engineering, chemical, and engineering, electrical & electronic—where female professors, even if noticeably underrepresented, have a higher average productivity than that of their male colleagues. For an in-depth analysis of all 145 SCs, the reader may refer to data provided in the Supplementary material.

**Table 9**
*Top ten SCs by gap of research performance (in terms of mean FSS) in favour of men, and the same for women, in Italy*

| | | F | | M | | |
|---|---|---|---|---|---|---|
| SC* | Discipline** | Obs. | FSS | Obs. | FSS | F vs M |
| Medical laboratory technology | 6 | 17 | 0.40 | 19 | 1.62 | -1.21 |
| Psychology, social | 8 | 50 | 0.65 | 32 | 1.72 | -1.07 |
| Information science & library science | 9 | 3 | 0.98 | 10 | 2.03 | -1.05 |
| Medicine, research & experimental | 6 | 28 | 0.51 | 38 | 1.41 | -0.90 |
| Computer science, interdisciplinary applications | 9 | 9 | 0.42 | 35 | 1.30 | -0.89 |
| Planning & development | 11 | 9 | 0.53 | 10 | 1.40 | -0.86 |
| Cell biology | 5 | 349 | 0.64 | 280 | 1.46 | -0.82 |
| Orthopaedics | 7 | 6 | 0.24 | 107 | 1.06 | -0.82 |
| Chemistry, medicinal | 6 | 280 | 0.72 | 221 | 1.40 | -0.69 |
| Haematology | 6 | 139 | 0.60 | 201 | 1.27 | -0.67 |
| … | | | | | | |
| Geography, physical | 4 | 8 | 1.08 | 26 | 0.90 | 0.18 |
| Physics, nuclear | 2 | 19 | 1.10 | 73 | 0.89 | 0.21 |
| Telecommunications | 9 | 22 | 1.24 | 135 | 1.00 | 0.24 |
| Social sciences, interdisciplinary | 10 | 27 | 1.30 | 25 | 1.04 | 0.26 |
| Materials science, biomaterials | 9 | 21 | 1.08 | 10 | 0.82 | 0.26 |
| Transportation | 11 | 13 | 1.46 | 66 | 0.95 | 0.51 |
| Health care sciences & services | 7 | 9 | 1.42 | 13 | 0.90 | 0.51 |
| Nursing | 7 | 19 | 1.24 | 6 | 0.61 | 0.63 |
| Psychology, applied | 8 | 12 | 1.59 | 11 | 0.81 | 0.78 |
| International relations | 10 | 3 | 1.33 | 3 | 0.28 | 1.05 |

\* SCs listed here are only those with at least three men and three women.
\*\* 1, Mathematics; 2, Physics; 3, Chemistry; 4, Earth and space sciences; 5, Biology; 6, Biomedical research; 7, Clinical medicine; 8, Psychology; 9, Engineering; 10, Political and social sciences; 11,

*Economics*

**Table 10**
*Top ten SCs by gap of research performance (in terms of mean FSS) in favour of women, and the same for men, in Norway*

| SC* | Discipline** | F Obs. | F FSS | M Obs. | M FSS | F vs M |
|---|---|---|---|---|---|---|
| Toxicology | 6 | 4 | 0.33 | 5 | 2.47 | -2.14 |
| Statistics & probability | 1 | 4 | 0.86 | 25 | 2.39 | -1.53 |
| Planning & development | 11 | 11 | 0.33 | 9 | 1.74 | -1.41 |
| Operations research & management science | 1 | 6 | 0.54 | 22 | 1.90 | -1.36 |
| Public administration | 11 | 7 | 0.17 | 5 | 1.51 | -1.34 |
| Substance abuse | 7 | 4 | 0.60 | 6 | 1.88 | -1.28 |
| Virology | 6 | 4 | 0.34 | 3 | 1.59 | -1.25 |
| Astronomy & astrophysics | 2 | 5 | 0.15 | 22 | 1.36 | -1.21 |
| Physics, applied | 2 | 4 | 0.52 | 19 | 1.66 | -1.15 |
| Cell biology | 5 | 15 | 0.35 | 19 | 1.42 | -1.07 |
| Radiology, nuclear medicine, & medical imaging | 6 | 3 | 0.87 | 14 | 0.47 | +0.40 |
| Sport sciences | 7 | 18 | 1.85 | 33 | 1.41 | +0.43 |
| Orthopaedics | 7 | 3 | 1.02 | 6 | 0.52 | +0.50 |
| Engineering, chemical | 9 | 5 | 1.52 | 29 | 1.02 | +0.50 |
| Geriatrics & gerontology | 7 | 8 | 1.01 | 5 | 0.48 | +0.53 |
| History & philosophy of science | 10 | 6 | 2.15 | 4 | 1.51 | +0.64 |
| Pharmacology & pharmacy | 6 | 21 | 1.04 | 11 | 0.24 | +0.80 |
| Communication | 10 | 11 | 1.66 | 23 | 0.82 | +0.84 |
| Engineering, electrical & electronic | 9 | 3 | 2.27 | 37 | 0.65 | +1.62 |
| Peripheral vascular disease | 7 | 3 | 2.03 | 3 | 0.33 | +1.70 |

\* The SCs listed here are only those with at least three men and three women.
\*\* 1, Mathematics; 2, Physics; 3, Chemistry; 4, Earth and space sciences; 5, Biology; 6, Biomedical research; 7, Clinical medicine; 8, Psychology; 9, Engineering; 10, Political and social sciences; 11, Economics

### 5.2.2 Gender performance differences among academic ranks

In this subsection, we analyse the average and median performance according to gender, disaggregating the data according to the academic rank of professors in the data set.

As indicated in Table 1, the breakdown of full, associate, and assistant professors between the two countries is very different. In Italy, professors in the data set are almost evenly distributed among the three ranks (although proportion differences occur across disciplines), with a slight prevalence of associate professors, representing 39% of the total. In Norway, the number of professors significantly increases with the rank, with a full professor to assistant professor ratio of 6:1.

Table 11 presents the average and median performance of professor subsets constructed combining gender, academic rank, and country. In Italy (columns 3–5), the performance gap between men and women decreases with rank, varying from a minimum of 0.15 for assistant professors to a maximum of 0.26 for full professors in terms of average (0.05–0.09 for medians). The gender productivity gap of Norwegian full professors is similar to that recorded for Italian full professors when measured with mean and median values. The gender gap for Norwegian assistant professors is much smaller than that for Italian ones (compared to the median it is not even significant). The case of Norwegian associate professors is peculiar: their average productivity is much lower than that recorded for assistant professors and there are no significant differences between the two genders.

Finally, the last two columns of Table 11 present the comparison between the two countries and reveal that for full professors, there are no significant differences for either gender. Instead, Italian associate professors are much more productive than their Norwegian colleagues of the same rank and gender, with a larger gap for men (0.52 in terms of mean values, 0.36 of median) than for women (0.35, 0.27). On the contrary, Norwegian assistant professors are, on average, more productive than their Italian colleagues of equal rank and gender, with a larger gap for women (-0.25 in terms of mean values, -0.17 of median) than for men (-0.17, -0.11).

*Table 11: Average and median normalized performance (FSS) of professors, by gender, academic rank, and country*

|  |  | Italy | | | Norway | | | Italy vs Norway | |
| --- | --- | --- | --- | --- | --- | --- | --- | --- | --- |
| Academic rank |  | M | F | $\Delta^\dagger$ | M | F | $\Delta^\dagger$ | $\Delta M^\dagger$ | $\Delta F^\dagger$ |
| Full professors | Avg. | 1.33 | 1.07 | 0.26 | 1.29 | 1.05 | 0.24 | 0.04 | 0.02 |
|  | Median | 0.73 | 0.64 | 0.09*** | 0.71 | 0.63 | 0.08* | 0.02 | 0.01 |
| Associate professors | Avg. | 1.09 | 0.91 | 0.17 | 0.57 | 0.56 | 0.00 | 0.52 | 0.35 |
|  | Median | 0.65 | 0.59 | 0.07*** | 0.29 | 0.31 | -0.03 | 0.36*** | 0.27*** |
| Assistant professors | Avg. | 0.72 | 0.57 | 0.15 | 0.90 | 0.82 | 0.07 | -0.17 | -0.25 |
|  | Median | 0.40 | 0.35 | 0.05*** | 0.51 | 0.52 | -0.02 | -0.11*** | -0.17*** |

*Two-sample Wilcoxon rank-sum (Mann-Whitney) test. Statistical significance: \*p-value <0.10, \*\*p-value <0.05, \*\*\*p-value <0.01*

$^\dagger$ *We show only two decimal digits for M and F values, but for Δ calculation we use the values without approximation.*

### 5.3 Gender representation vs research performance

As reported in the literature (Section 2), it is likely that different confounding factors determine the current underrepresentation of women, overall and within each academic rank, in both Italy and Norway. In this section, we attempt to answer the following question: if the academic rank were assigned on the basis of research performance in the five-year period under observation, would the current gender representation remain the same?

To answer that question, we divide the ranking of professors in each country into three classes. The size of each class corresponds to that of each academic rank in the country. For example, if full professors in a country are 25% of overall faculty, the first class by performance is quartile 1. Then, we measure the gender proportions in each class of performance (we call it expected representation) and compare them with current representation. Any differences would denote underrepresentation by either gender.

For reasons of significance, we have excluded from the analysis those SCs with less than four professors in each country. The restriction causes the Norway population under observation to reduce slightly from 4,214 to 4,195 professors, while remaining the same for Italy. The results of the exercise are presented in Table 12. We observe that in both Italy (-9.0%) and Norway (-6.5%), female full professors would appear underrepresented if academic rank were aligned with research performance. In Italy, slight underrepresentation also occurs for female associate professors (-0.8%).

The overrepresentation of women in the lowest rank is also evident: the difference between the current and expected proportion of female assistant professors is +9.9% for Italy and +6.3% for Norway.

From the above analysis, we cannot conclude that women suffered discrimination, because we do not know their comparative merits at the time of career advancement. While the study is based on a five-year period, it may take a much longer time to qualify or obtain the position of full professor. Assumedly, relatively speaking, there are a greater number of younger female professors with shorter careers.[4] Moreover, parameters other than scientific publishing also matter in these processes. Nevertheless, the findings raise a question concerning the fairness of the career system from a gender perspective, which would be interesting to analyse in further studies. In fact, it is difficult to find good reasons why women would present a lower performance than men at the time of competition for recruitment and career progress (which would legitimize the current gender unbalance across academic ranks), but improve their performance relative to men thereafter.

Table 12
*Current and expected representation of professors, by academic rank, gender, and country*

| Academic rank | Country | Current | | | | Expected | | | | F/Total Current vs Expected |
|---|---|---|---|---|---|---|---|---|---|---|
| | | F | M | Total* | F/total | F | M | Total* | F/total | |
| Full professor | Italy | 1780 | 7936 | 9716 | 18.3% | 2662 | 7079 | 9741 | 27.3% | -9.0% |
| | Norway | 645 | 1825 | 2470 | 26.1% | 802 | 1660 | 2462 | 32.6% | -6.5% |
| Associate professor | Italy | 4375 | 8066 | 12441 | 35.2% | 4456 | 7927 | 12383 | 36.0% | -0.8% |
| | Norway | 583 | 676 | 1259 | 46.3% | 440 | 783 | 1223 | 36.0% | +10.3% |
| Assistant professor | Italy | 4662 | 5222 | 9884 | 47.2% | 3699 | 6218 | 9917 | 37.3% | +9.9% |
| | Norway | 193 | 273 | 466 | 41.4% | 179 | 331 | 510 | 35.1% | +6.3% |

*\* SCs with at least four professors, in each country*

## 6. Discussion and conclusions

Comparing the occasionally contradicting results of studies on the gender performance gap is not as easy as it may appear. In fact, in addition to contexts, fields, and methods, the bibliometric indicators used to assess performance are rather diverse. A few scholars recurred to (full or fractional) publication counting, others to average citations per paper, others more to the h-index. The fields of observation are heterogeneous too, as some are country-level analyses, others do not distinguish across countries, a few are specific to institutions or a research field, while others do not distinguish across fields.

Our study has contributed to further knowledge on research performance by gender in Italy and Norway. It is evident that women are significantly underrepresented in the academic staff, accounting for approximately one-third of the population analysed in both countries. However, the proportion varies significantly across fields and academic positions, thereby reflecting a vertical and horizontal segregation. This segregation is apparently rather universal and extends to most other countries as well (European Commission 2019).

We find large differences in the performance of men and women measured by bibliometric methods. The average FSS score of men is 37% higher than that for women in Italy and 32% higher in Norway. This gender gap primarily relates to the volume of the scientific production, while the differences in the other parameters (citation impact and journal profile) are small. In conclusion, not only are women underrepresented, but their average overall research performance is also inferior compared with men. While these findings correspond with numerous previous studies, both issues are reasons for serious concerns considering the role of women in science.

---

[4] At 31/12/2015, the mean age of male and female Italian professors in the data set was, respectively, 53 and 51 years.

Further, the gender performance differences are present at the level of disciplines and academic positions. It appears that, on average, men outperform women in both countries. Actually, the median performance differences between genders are statistically significant only in Italy in all disciplines but political and social sciences. In Norway, the statistical significance of median performance differences is often missing at the discipline level. Possible reasons for this could be the lower number of observations in Norway, which reduces the statistical power, as well as the fact that the gender performance gap in Norway is smaller than that in Italy.

Further, in both countries, full professors achieve the highest performance and this holds true for both genders. However, the gender gap is also largest for this staff group. For the other two ranks, the analysis reveals major differences between the two countries. While the average performance of Italian professors is aligned with their academic ranks (associate professors outperform assistant professors), the opposite holds true in Norway, whereby assistant professors achieve higher performances than associate professors. These patterns hold true for both genders. It is possible that the deviating findings may be influenced by the different academic career systems of the two countries. In Norway, associate professors are entitled to apply for promotion to full professor on the basis of research competence. Thus, one does not need to be employed in open full professorships posts. Consequently, the proportion of full professors is higher in Norway than in numerous other countries (Frølich et al., 2018). Further, the career selection mechanisms in Italy may be stronger than those in Norway. The deviating findings for Norwegian associate professors may possibly be explained by an accumulation of individuals with few research activities and who accordingly are not able to qualify for full professorships. However, we do not have data available to verify this.

Scientific performance is rather skewed at the level of individuals. This implies that indicators based on arithmetic mean are strongly influenced by the presence of high-performing researchers (this is why in our work we also offered median values and non-parametric tests of significance of differences). Nevertheless, the underlying distribution has been seldom analysed and most studies apply the mean as an overall measure, including numerous previous reports on gender performance differences (e.g. European Commission, 2019). In line with a few previous studies (Bordons, Morillo, Fernández, & Gómez, 2003; Abramo, D'Angelo, & Caprasecca, 2009b), our results reveal that most of the overall gender differences in both countries can be explained by the tails of the distributions—particularly a much higher proportion of men among the top performing scientists. Findings confirm previous results by Abramo, D'Angelo, and Caprasecca (2009b), who found a remarkable gender performance gap among the top 10% performers; in contrast, in the remaining 90% of the population, differences even reverse in favour of women in the case of full professors. This suggests that oft-heard generalizations, like 'women have lower performance than men' are incorrect. As a matter of fact, the performances of men and women performances do not differ much, expect in the top performing groups.

In addition to the empirical results provided, the value added of our study is related to the method and indicator of performance measurement, which includes several aspects of the research performance and, most importantly—differently from other available indicators—it accounts for inputs (i.e. the cost of research). Because women are relatively more underrepresented in the higher academic ranks, characterized by comparatively higher salaries, not accounting for costs does indeed favour men.

Although the FSS is not affected by such bias, it is not entirely immune from it. In fact, because of the lack of data, the FSS does not account for absence like maternity and sick leaves and, in general, for the actual hours devoted to research. It is assumed that the yearly average cost of labour and time devoted to research are the same for men and women. While the cost of labour does not depend on gender in both countries, the assumption that time for research is invariant too does not reflect reality, at least in Italy.

It is well-known that the new family trends and patterns have been paralleled by changes in gender roles, particularly an expansion of the female role to an economic provider for a family. However, female emancipation has not progressed at the same pace across countries.

As discussed in the introduction, the societal role of women in Italy and Norway differs and women in Italy have retained more of their traditional engagements than in Norway. Extensive involvement in family responsibilities may hamper an academic career. It is possible that these societal differences may partially explain why the gender performance gap is larger in Italy than in Norway.

As noted above, the performance gap among top performing scientists contributes in a preponderant manner to determining the gap for the entire Italian and Norwegian populations of professors. It is evident that in order to achieve levels of output that are the same as those of top scientists, an extraordinary amount of time on research is required at the expense of other things. Because of different societal roles, men are perhaps more willing, or able, to prioritize in this manner than women.

Further, a few of the causes that determine the gender gap can be reduced through the formulation of opportune policies and specific initiatives. There is currently much attention paid to the problem in numerous countries. In Norway, the Research Council has funded a program to support efforts of the research institutions to promote gender balance.[5] This includes actions such as mentor programs, network programs, and scholarships directed towards female researchers. Nevertheless, such policies will only achieve their effects over the long term. Therefore, it is necessary to simultaneously develop and adopt performance evaluation systems where female researchers are evaluated equitably with men. The adoption of such systems becomes more urgent with the increasing reliance on bibliometric indicators in supporting decisions related to academic recruitment and advancement as well as in the granting of project funds (Fischer, Ritchie, & Hanspach, 2012); other factors are also emphasized in these contexts. The setting and role of bibliometric indicators may vary, but an interesting question is whether the use of such indicators will contribute to a fairer evaluation of female researchers.

On the one hand, the use of bibliometric indicators and the gender differences in productivity at a general level may hamper the career of women, as they may lose in competition with men—although the differences mainly relate to the top performers, as described above. On the other hand, the use of bibliometric indicators might enable female researchers to obtain more equitable evaluation compared to promotion systems based on qualitative judgment by a committee of experts, which is typical of numerous university systems. In fact, such experts are generally selected among academics with the highest academic ranks, where men are still clearly dominant (Moss-Racusin, Dovidio, Brescoll, Graham, & Handelsman, 2012; Lariviere, Ni, Gingras, Cronin, & Sugimoto, 2013) and where the purely qualitative committee judgment could then be influenced by subconscious gender bias (Sheltzer & Smith, 2014).

---

[5] https://www.forskningsradet.no/en/apply-for-funding/funding-from-the-research-council/balanse/. Accessed 16 February 2021.

Due to their quantitative nature, bibliometric indicators can reduce these effects, but the conditions for a fair evaluation can only be met through the use of methods and indicators that do not implicitly disfavour female researchers (O'Brien & Hapgood, 2012). The literature contains a small number of studies that investigate this specific theme, offering solutions for modifying traditional bibliometric indicators and/or evaluation methods that account for the gender of researchers.

Symonds, Gemmell, Braisher, Gorringe, and Elgar (2006) recommend opting for an evaluation that considers not only the number of publications, which would penalize female researchers, but also their impact. As a method, they propose beginning with the h-index, on which they perform a series of corrections. In line with this study, Cameron, White, and Gray (2016) also suggest modifications to the h-index, operating two changes considered capable of eliminating the bias that implicitly favours male researchers. First, they propose using the h-index adjusted for career age, excluding the inactive years of the female researchers due to periods of maternity. Second, the authors suggest eliminating self-citation from the calculation of the h-index, as male researchers—being characterized by greater scientific self-confidence—tend to self-cite more than women (Nielsen, 2016b).

Adopting a more radical approach, Abramo, Cicero, and D'Angelo (2015) advance the proposal of constructing performance rankings distinguished by gender, where the context renders this justifiable. They indicated that in Italy "the comparison of individuals' research performance, between peers of the same gender, leads to rank positions that are detectably different compared to those from ranking lists constructed without distinction by gender: in roughly 70% of the individual disciplines analysed, women professors obtain a shift ahead in the ranking lists with distinction by gender."

Before opting for the above solution, it appears logical that the decision-maker must at least verify the extent of the gender gap in performance and explore the possible causes. It would also be desirable that the concerned parties—male and female researchers—be called on for their opinions and to avoid negative reactions to a possible separation of rankings.

**Acknowledgement**
This research was partly funded by the Research Council of Norway (grant number 295817).